\newif
\ifconfver
\confvertrue        

\ifconfver
    \documentclass[10pt,journal,final,twoside]{IEEEtran}
\else
    \documentclass[11pt,draftcls,onecolumn]{IEEEtran}
\fi

\usepackage{amsthm}
\usepackage{amsmath}
\usepackage{fixmath}
\usepackage{amsfonts}

\ifCLASSOPTIONcompsoc
  \usepackage[nocompress]{cite}
\else
  \usepackage{cite}
\fi
\usepackage{graphicx}
\usepackage{algorithm}
\usepackage{algorithmic}
\usepackage{accents}
\usepackage{array,color}
\usepackage{subfig}
\usepackage{dsfont}

\def\v{{\mathbf v}}

\def\V{{\mathbf V}}
\def\u{{\mathbf U}}
\def\u{{\mathbf u}}

\newcounter{proposition} 
\newenvironment{proposition}{ \vspace{0.1cm} \noindent {\bf Proposition
\theproposition.} \addtocounter{proposition}{-1}\refstepcounter{proposition}\em } { \addtocounter{proposition}{1}
\vspace{0.1cm} \normalfont }

\newcounter{remark} 
\newenvironment{remark}{\vspace{0.1cm} \noindent {\bf Remark \theremark.} \addtocounter{remark}{-1}\refstepcounter{remark}} {\addtocounter{remark}{1}
	\vspace{0.1cm} \normalfont }
    
\begin{document}

\title{Fast Nonconvex SDP Solvers for Large-scale Power System State Estimation}

\author{Yu~Lan,
        Hao~Zhu,
        and~Xiaohong~Guan
\IEEEcompsocitemizethanks{\IEEEcompsocthanksitem This work was supported in part by the National Science Foundation under Award ECCS-1802319 and the Key R\&D Project of China 2016YFB0901900. 
\IEEEcompsocthanksitem Y. Lan is with Ministry of Education Key Lab for Intelligent Networks and Network Security, Xi'an Jiaotong University, Xi'an, Shaanxi, 710049, China. Y. Lan is also with the Department of ECE, The University of Texas at Austin, Austin, TX 78705, USA. E-mail: ylan@sei.xjtu.edu.cn
\IEEEcompsocthanksitem H. Zhu is with the Department of ECE, The University of Texas at Austin, Austin, TX 78705, USA. Email: haozhu@utexas.edu
\IEEEcompsocthanksitem X. Guan is with Ministry of Education Key Lab for Intelligent Networks and Network Security, Xi'an Jiaotong University, Xi'an, Shaanxi, 710049, China. E-mail: xhguan@xjtu.edu.cn
}}

\markboth{}%
{Lan \MakeLowercase{\textit{et al.}}: Fast Nonconvex SDP Solvers for Large-scale Power System State Estimation}

\IEEEtitleabstractindextext{%
\begin{abstract}
Fast power system state estimation (SE) solution is  of paramount importance for achieving real-time decision making in power grid operations. Semidefinite programming (SDP) reformulation has been shown effective to obtain the global optimum for the nonlinear SE problem, while suffering from high computational complexity. Thus, we leverage the recent advances in nonconvex SDP approach that allows for the simple first-order gradient-descent (GD) updates.  Using the power system model, we can verify that the SE objective function enjoys nice properties (strongly convex, smoothness) which in turn guarantee a linear convergence rate of the proposed GD-based SE method. To further accelerate the convergence speed, we consider the accelerated gradient descent (AGD) extension, as well as their robust versions under outlier data and a hybrid GD-based SE approach with additional synchrophasor measurements. 
Numerical tests on the IEEE 118-bus, 300-bus and the synthetic ACTIVSg2000-bus systems have demonstrated that FGD-SE and  AGD-SE, can approach the near-optimal performance of the SDP-SE solution at  significantly improved computational efficiency, especially so for AGD-SE.
\end{abstract}}


\maketitle

\IEEEdisplaynontitleabstractindextext

\IEEEpeerreviewmaketitle

\section{Introduction}
\label{intro}

\IEEEPARstart{P}{ower} system state estimation (SE) aims to obtain the operating condition of the grid, namely nodal complex voltages, from noisy measurements at buses and branches. The SE problem is of paramount importance for reliable control and economic operation of power systems; see e.g., \cite{wood2012power},\cite{giannakis2013monitoring}. 

Due to a nonlinear measurement model, SE is traditionally formulated as a nonlinear least-squares (LS) problem and solved by Gauss-Newton (GN) iterations \cite{monticelli2000electric}. The GN method iteratively updates the variables by minimizing an approximate objective through linearization. Albeit computationally efficient  at each iteration, convergence of GN to global optimum is generally not guaranteed. As shown by recent work \cite{zhao2018statistical}, heavy grid loading conditions or bad data such as topology errors can easily lead to divergent GN iteration. To tackle this, a semidefinite programming (SDP) reformulation of the SE problem has been proposed in \cite{zhu2014power} using rank relaxation. To promote lower-rank solutions, \cite{weng2015convexification} suggested a nuclear norm based penalization. General penalization terms are designed in \cite{zhang2017conic,madani2016convexification} for guaranteed exact recovery and quantifiable estimation error of the SDP-SE formulation. 
To account for bad data, robust SDP-SE formulation is available by modeling outliers using a sparse vector \cite{kekatos2017psse, madani2017power}. 
Although the global optimum to SDP-SE can be obtained by generic algorithms such as the interior-point method, the high-order polynomial complexity therein could greatly challenge real-time implementation in large-scale systems \cite{kekatos2017psse}. Recent work has focused on using conic relaxation or composite optimization techniques for the SE problem \cite{wang2019robust1,zhang2017conic}. A parallelizable SDP solution was also developed in \cite{zhu2014power,madani2017low} using graph-specific decomposition. However, it remains open to develop fast solvers at simple implementation steps for the SDP-SE formulation.

Recently, a nonconvex approach to solving SDPs by representing the solution matrix using the Burer-Monteiro factorization \cite{burer2003nonlinear, burer2005local} has become popular. The factorization form can easily to eliminate the positive semi-definite (PSD) constraint of SDP problems. 
This way, the first-order gradient descent (GD) updates are readily applicable to the resultant nonconvex objective, leading to  the so-termed factored GD (FGD) algorithm  \cite{bhojanapalli2016dropping}. Linear convergence guarantee for FGD is available if the original SDP objective is strongly convex and smooth with sufficiently close initialization. 
For ill-conditioned objective, the FGD may suffer from slow convergence speed. To tackle this, the popular accelerated GD (AGD) method by Nesterov \cite{nesterov1983method} has been extended to solve the nonconvex SDP reformulation \cite{kyrillidis2018run}. 

Our goal is to accelerate power system SE by leveraging the latest advances in fast nonconvex SDP solution techniques. Towards this end, we first reformulate the SDP-SE problem using the matrix factorization approach. The FGD-based SE method is developed, along with the objective function analysis for establishing the linear convergence rate conditions. Moreover, the AGD-SE method is designed  for accelerated convergence speed.  Furthermore, to account for practical concerns, we have developed the robust FGD-/AGD-SE to address outliers via a simple hard thresholding operation, as well as accounted for the additional synchrophasor measurements for a hybrid GD-based SE methods. All the GD-SE methods and extensions enjoy extremely efficient update per iteration, with linear convergence rate verified by our objective function analysis and extensive numerical tests. Numerical results also confirm the near-optimal performance of the GD-SE methods, in estimating the voltage phasors and identifying outlier meters.  



The rest of the paper is organized as follows. The SDP-SE problem and the nonconvex reformulation are introduced in Section \ref{section2}. Section \ref{section3} presents the FGD-/AGD-SE methods along with the convergence analysis based on the power system models. Section \ref{section4} develops the robust GD-SE methods and the PMU-aided extensions. Several numerical tests presented in Section \ref{section5} corroborate the faster computation time of FGD and AGD relative to the SDP-SE solver, and improved estimation performance over GN iterations. The paper is wrapped up in Section \ref{section6}.

\emph{Notation:} Upper (lower) boldface symbols stand for matrices (vectors); $|\cdot|$ stands for the magnitude;  $(\cdot)^\mathcal{T}$ denotes transposition; $(\cdot)^{\mathcal{H}}$ complex-conjugate transposition; $\Re (\cdot)$/$\Im (\cdot)$ the real/imaginary part; $\mathrm{Tr}(\cdot)$ the matrix trace; {rank}$(\cdot)$ the matrix rank; $\|\cdot\|_F$ the Frobenius norm; $\|\cdot\|_2$ the spectral norm, and $\mathbf{V} \succeq \mathbf{0}$ denotes $\mathbf{V}$ is a positive semi-definite (PSD) matrix.







\section{Problem Formulation}
\label{section2}

Consider a transmission network modeled as a graph $\mathcal{G}=(\mathcal{N},\mathcal{E})$, with the set of buses (nodes) in $\mathcal{N} := \{1,...,N\}$ and the set of lines (edges) in $\mathcal{E} :=\{(n,n')\}$. The complex voltage phasor $V_n$ per bus $n\in \mathcal{N}$ can be expressed in the rectangular coordinate as $V_n=\Re(V_n)+\mathtt{j}\Im(V_n)$. All nodal voltages form the full system state vector $\mathbf{v}:=[V_1,...,V_N]^{\mathcal{T}}\in\mathbb{C}^N$. To estimate nodal voltages in $\mathbf{v}$, a subset of the following system variables are measured: 
\begin{itemize}
\item $P_n (Q_n)$: the active (reactive) power injection at bus $n$;
\item $P_{n n'}(Q_{n n'})$: the active (reactive) power line flow from bus $n$ to bus $n'$;
\item $\left| {{V_n}} \right|$: the voltage magnitude at bus $n$.
\end{itemize}
The ac power flow model \cite[Ch. 4]{wood2012power} asserts power variables are nonlinearly (quadratically) related to the state $\mathbf{v}$. 

Collecting the noisy measurements in the vector $\mathbf{z} :=\big[\{\check{P}_n\}
, \{\check{Q}_n\}
, \{\check{P}_{nn'}\}
, \{\check{Q}_{nn'}\}
,\{|\check{V}_n|^2\}
\big]^T\in\mathbb{R}^L$, where $L$ is the total number of measurements, one can write the $\ell$-th measurement in $\mathbf{z}$ as
\begin{align}\label{eq:measureh}
z_\ell=h_\ell(\mathbf{v})+\epsilon_\ell,~\forall \ell = 1,\ldots, L
\end{align}
where $h_\ell(\cdot)$ stands for the nonlinear transformation from $\mathbf{v}$, and $\epsilon_\ell$ accounts for the additive measurement error. 
Given this measurement model, the SE problem can be cast as a (weighted) LS minimization one over $\v$. 
Without loss of generality (Wlog), we assume the weight coefficients are included by the model \eqref{eq:measureh}. Thus, it suffices to consider the \textit{unweighted} LS-SE objective throughout the paper. Due to its nonlinear objective, Gauss-Newton (GN) has been the workhorse solution for LS-SE; see e.g., \cite{gomez2004power}. GN iteratively approximates the objective by linearizing \eqref{eq:measureh} at the latest solution. This iterative linearization procedure, though computationally efficient if convergent, can be potentially divergent under heavy loading or bad data conditions; see e.g., \cite{zhao2018statistical,zhang2018spurious}.

One approach to tackle the nonlinearity is to introduce the outer-product matrix $\mathbf{V}:=\mathbf{v}\mathbf{v}^{\mathcal{H}} \in\mathbb{C}^{N\times N}$, consisting of all quadratic terms involving $\mathbf{v}$. 
This way, each measurement in (\ref{eq:measureh}) is linearly related to $\V$, as given by 
\begin{align} \label{eq:noisy}
z_\ell=\v^\mathcal{H}\mathbf{H}_\ell \v +\epsilon_\ell =\mathrm{Tr}(\mathbf{H}_\ell \mathbf{V})+\epsilon_\ell,~\forall \ell = 1,\ldots, L.
\end{align}
where $\mathbf{H}_\ell \in\mathbb{C}^{N\times N}$ is Hermitian matrix depending on the network topology and line parameters (see e.g., \cite{zhu2014power,wang2019robust2} for the definitions). Reformulating the LS objective about $\v$ using \eqref{eq:noisy} leads to the following semidefinite program (SDP) for $\V$:
\hspace*{-6pt}
\begin{subequations}\label{eq:SDP} 
\begin{align}
{\hat{\mathbf V}}= \arg\min_{\mathbf{V} 
} &~ f(\mathbf{V}):= \sum_{\ell=1}^L \frac{1}{2}\left[z_\ell -\mathrm{Tr}(\mathbf{H}_\ell \mathbf{V})\right]^2 \label{eq:SDPo}\\
\mathrm{s.t.}\:&\mathbf{V} \succeq \mathbf{0} \label{eq:SDPc}
\end{align} 
\end{subequations}
where the PSD constraint in \eqref{eq:SDPc} together with {rank}$(\mathbf{V})=1$ can guarantee the existence of $\mathbf{v}$ that satisfies $\hat{\mathbf{V}}=\mathbf{v}\mathbf{v}^{\mathcal{H}}$.
Due to the nonconvexity of rank constraint, it is dropped through a well-appreciated semidefinite relaxation (SDR) procedure that leads to the convex SDP-SE formulation \eqref{eq:SDP}. Solving the latter can achieve a near-optimal performance for the SE problem as the solution tends to have very low-rank; see \cite{kekatos2017psse,madani2016convexification}. General convex solvers such as the popular interior-point method based solver SeDuMi \cite{sturm1999using} can obtain the optimal $\hat{\mathbf V}$ in polynomial time. Nonetheless, these solution methods can scale unfavorably as the number of buses or measurements increase, with worst-case complexity at $\mathcal O (N^{4.5})$ \cite{kekatos2017psse}. Thus, it is necessary to develop accelerated algorithms for solving large-scale SDP-SE in real-time.

Motivated by recent work on nonconvex SDP solvers \cite{bhojanapalli2016dropping}, we consider an equivalent formulation of \eqref{eq:SDP}. The idea is to use matrix factorization to represent the PSD matrix as product of two factor matrices, as first proposed by \cite{burer2003nonlinear,burer2005local}. Accordingly, it can tackle the main computational challenge caused by the PSD conic constraint \eqref{eq:SDPc}.
Expressing $\mathbf{V}=\mathbf{u}\mathbf{u}^\mathcal{H}$ with vector $\mathbf{u}\in \mathbb C^{N}$, one can reformulate \eqref{eq:SDP} as an unconstrained one involving the complex $\mathbf{u}$, namely, 
\begin{align}\label{eq:gU}
\hat{\mathbf u} = \arg \min_{\mathbf{u}\in \mathbb{C}^{N}}\:g(\mathbf{u}):=f(\mathbf{u}\mathbf{u}^\mathcal{H}).
\end{align}
In general, we can use a rank-$r$ ($r\geq 1$) matrix component. The advantage of using a rank-one $\mathbf{u}$ here is two-fold: i) the relaxed SDP problem \eqref{eq:SDP} is likely to attain a nearly rank-one solution; and ii) searching for rank-one solutions has the lowest computational complexity. Although the LS objective $f$ is convex in $\mathbf V$, $g(\mathbf u)$ in \eqref{eq:gU} becomes nonconvex again. Interestingly, this reformulation is equivalent to the LS-SE objective. Nonetheless, as opposed to the GN method, we will present gradient descent (GD) solutions that can achieve guaranteed recovery performance. 

\section{Gradient-descent Based SE Solvers}
\label{section3}
Thanks to the unconstrained structure and convenient gradient computation, we can use first-order methods such as gradient descent (GD)  to solve \eqref{eq:gU}. This simple approach, termed as factored GD (FGD) in \cite{bhojanapalli2016dropping}, has been shown to converge to the global optimum of general SDP problems, at a linear rate similar to that of classical GD method for strongly convex and smooth functions \cite[Ch. 9]{boyd2004convex}. Inspired by these results, we first use power flow analysis to prove the convergence rate of FGD for our LS-SE problem. Furthermore, we will propose an accelerated scheme of FGD for improved convergence rate. 

\subsection{Factored Gradient Descent Method}
\label{subsecFGD}

To develop the FGD updates for \eqref{eq:gU}, it suffices to compute the derivative $\nabla g(\mathbf{u})$ for any complex $\mathbf{u}$. Using the chain rule, one can write
\begin{align}\label{eq:gradient}
\nabla g(\mathbf{u})= 2 \nabla f(\u\u^\mathcal H) \times \u = 
\sum_{\ell=1}^L 2 \big[\mathbf{u}^\mathcal{H} \mathbf{H}_\ell \mathbf{u} -z_\ell\big] \mathbf{H}_\ell \mathbf{u}.
\end{align}
The rigorous derivation for a complex derivative is slightly more complicated, with the details in Appendix \ref{Appendix:gradient}. Given step-size $\eta$, 
%
the main FGD-SE steps are tabulated in Algorithm \ref{alg:FGD} for general rank-$r$ solutions. To select the $\eta$ value and establish the convergence results, one needs to investigate the characteristics of the original SDP function $f(\mathbf V)$.

Specifically, consider a compact form 
$f(\mathbf{V})=\frac{1}{2}\|\mathbf{z}-\mathcal{H}(\mathbf{V})\|_2^2$, 
where $\mathcal{H}(\cdot)$ stands for the linear mapping in the SDP objective \eqref{eq:SDP}. In general, strong convexity is needed for the loss function $f$. Nonetheless, it is challenging to show that over any PSD $\mathbf V$ of an arbitrary rank $r$. Instead, we can restrict it to a subset of $\mathbf V$'s \cite{negahban2012restricted}. Specifically, we consider the subset  
\begin{align}\label{eq:setV}
    \mathcal{V}:=\{\V|\mathrm{rank}(\V)=1~ \mathrm{and}~\underaccent{\bar}{V}^2 \leq \V_{nn} \leq \bar{V}^2 ~\forall n\}
\end{align} where $\underaccent{\bar}{V}/\bar{V}$ are lower/upper bounds on bus voltage magnitude. These bounds (e.g., $\pm 5\%$ from unity) are easily available thanks to well-designed power system voltage control mechanism. Using the subset $\mathcal V$, we can show the following bounds as given in 
Appendix \ref{Appendix:m and M}.

\begin{proposition} \label{prop:m}
For any $\mathbf V \in \mathcal V$ as given by \eqref{eq:setV}, its mapped output  $\mathcal H(\mathbf V)$ for the SDP-SE objective $f$ in \eqref{eq:SDP} satisfies 
\begin{align}
m\cdot \|\mathbf{V}\|_F^2 \leq
\|\mathcal{H}(\mathbf{V})\|_2^2 \leq M\cdot \|\mathbf{V}\|_F^2 \label{eq:boundH}
\end{align}
where $m$ and $M$ are positive coefficients. Hence, the objective $f$ is $m$-strongly convex and $M$-smooth over restricted set $\mathcal V$ with the condition number $\kappa= {M}/{m}$. 
\end{proposition}


%
%

Proposition \ref{prop:m} is shown by leveraging the power flow model embedded in the mapping $\mathcal{H}$. It is a key result to ensure the convergence of FGD for the objective $g$, even though the latter is not (strongly) convex itself. As developed in \cite{bhojanapalli2016dropping}, using the initial $\V_0 = \u_0 \u_0^\mathcal{H}$ one can set the step-size according to
\begin{align}\label{eq:eta}
\eta=\dfrac{1}{16 (M\|\mathbf{V}_0\|_2+\|\nabla f(\mathbf{V}_0)\|_2)}
\end{align}
where the constant ratio $1/16$ is a rough number that can be tuned up based on specific problems.  In practice, the  smoothness parameter $M$ can also be approximated by $\|\nabla f(\mathbf{V}_0)-\nabla f(\mathbf{V})\|_F / \|\mathbf{V}_0-\mathbf{V}\|_F$ for any $\V \in \mathcal V$. 

\begin{algorithm}[!t]
\caption{FGD-SE}
\label{alg:FGD}
\begin{algorithmic}[1]
\REQUIRE  Function $f$, rank $r$, maximum iteration number $K$.
\ENSURE  $\mathbf{u}$ and $\mathbf{V}=\mathbf{u}\mathbf{u}^\mathcal{H}$
\STATE   Initialize $\mathbf{u}_0\in \mathbb{C}^{n\times r}$ and set $\mathbf{V}_0=\mathbf{u}_0\mathbf{u}_0^\mathcal{H}$. 
\STATE Set the step-size $\eta$ as in \eqref{eq:eta}. 
\FOR{$k=0$ to $K$}
\STATE  $\mathbf{u}_{k+1}=\mathbf{u}_k-\eta \nabla g(\mathbf{u}_k) $
\ENDFOR 
\RETURN $\mathbf{u} = \u_K$ and $\mathbf{V}=\mathbf{u}_K\mathbf{u}_K^\mathcal{H}$
\end{algorithmic}
\end{algorithm}


To establish the local convergence of Algorithm \ref{alg:FGD} for the SDP-SE problem, we need the following two assumptions regarding the approximation to the global optimal $\hat{\mathbf u}$ and $\hat{\mathbf V}^1:=\hat{\mathbf u}\hat{\mathbf u}^\mathcal H$, using small constant numbers $\rho_u$ and $\rho_v$.

\begin{itemize}
    \item[\bf (as1)] The initial $\u_0$ satisfies $\mathrm{Dist}(\u_0,\hat{\u}) \leq \frac{\rho_u}{\kappa} \|\hat{\u}\|_2 $, where $\mathrm{Dist}$ 
    denotes the minimum Euclidean distance between the two complex vectors up to any rotational change. 
    
    \item[\bf (as2)] The optimum $\hat{\V}$ satisfies $\|\hat{\V}-\hat{\V}^1\|_F \leq \frac{\rho_v}{\kappa^{1.5}} \|\hat{\u}\|_2^2 $.
\end{itemize}
Assumption (as1) requires the initial guess to be sufficiently close to the optimal $\hat{\mathbf u}$. As the upper bound scales with $\|\hat{\u}\|_2$, it could be reasonable for good initialization such as flat start or the SE solution from dc power flow model. Assumption (as2) relates to the rank-one approximation to $\hat{\V}$. Since the SDP-SE solution  $\hat{\V}$ is nearly rank-1 \cite{madani2016convexification,kekatos2017psse}, the upper bound therein could also be satisfied in practice.

\begin{proposition}
\label{prop:fgd_strongly_convergence}
Suppose (as1) and (as2) hold and use the step-size in \eqref{eq:eta}. Under Proposition \ref{prop:m} with the SE objective $f$ satisfying \eqref{eq:boundH}, the FGD-SE updates in Algorithm \ref{alg:FGD}  converge linearly to a neighborhood of $\hat{\u}$. Specifically, it can be shown that the update at iteration $k$ satisfies 
\begin{align} \label{eq:conv}
    \mathrm{Dist}(\u_{k+1},\hat{\u})^2 \leq \alpha\cdot \mathrm{Dist}(\u_k,\hat{\u})^2+\beta \cdot \|\hat{\V}-\hat{\V}^1\|_F^2
\end{align}
where $0 < \alpha < 1$ is the contraction rate and $\beta$ is a constant number, both depending on the optimal $\hat{\mathbf u}$.  
\end{proposition}

The proof directly follows from \cite{bhojanapalli2016dropping} based on the analysis of the iterative update. Note that \eqref{eq:conv} shows that the accuracy of the convergent solution depends on the term $\beta\|\hat{\V}-\hat{\V}^1\|_F^2$. If the optimum $\hat{\V}$ is rank-one, the FGD-SE updates will accordingly converge to the globally optimal $\hat{\V}$. This condition relates to the approximation performance of SDP-SE \eqref{eq:SDP} to the original unrelaxed problem, which has been corroborated by both analytical and numerical results \cite{madani2016convexification,kekatos2017psse,zhu2014power}. Hence, we will not focus on this aspect but instead discuss more on the convergence speed of FGD-SE, as related to the $\alpha$ value. Faster convergence speed requires a smaller $\alpha$, which critically depends on the initial $\u_0$ and the condition number $\kappa$.
Various initialization schemes are available for SE, including flat start and dc power flow based solution. Our numerical tests suggest that these options work well for the FGD-SE updates, even though they do not strictly satisfy (as1). %
Our empirical experience has identified the main issue to be that the SE objective $f$ tends to have a very large condition number $\kappa$. As a result, the convergence speed for FGD would gradually decrease as the number of iterations $k$ increases. This motivates us to develop an accelerated scheme for FGD-SE as follows.

\subsection{Accelerated Gradient Descent Method}

Slower convergence speed as a result of ill-conditioned objective function is a common issue for first-order methods. One popular improvement is to use the Nesterov's acceleration scheme \cite{nesterov1983method}. This improved first-order method has been shown to achieve superlinear convergence rate for convex objectives. Loosely speaking, it can reduce the condition number to $\sqrt{\kappa}$. Here, we develop the accelerated GD (AGD) based SE method as a heuristic alternative for the nonconvex problem \eqref{eq:gU}. 

Different from FGD that uses only the instantaneous gradient, AGD takes the information from the past two iterations to compute the update. Per iteration $k$, a time-varying interpolation is first performed to obtain 
\begin{align} \label{eq:AGD predict}
\mathbf{u}^+=\mathbf{u}_k+\left(\frac{k-2}{k+1}\right)(\mathbf{u}_k-\mathbf{u}_{k-1}), \forall k = 1, 2, \ldots
\end{align}
which is used to update $\u_{k+1}$ as the gradient descent on $\u^+$. The ratio $\mu= \frac{k-2}{k+1}$ is termed as the momentum parameter for AGD, which goes to 1 as $k$ increases. Hence, it is also possible to use a large constant $\mu <1$ for updating  $\mathbf{u}^+$.
%
For $\u_0$ at $k=0$, the update $\u_1$ is simply computed using the FGD rule. The AGD method is tabulated in Algorithm \ref{alg:AGD}. 

\begin{remark}
{\emph (AGD for nonconvex $g$.)} Although the convergence of AGD is well understood for convex functions \cite{nesterov1983method}, it is generally an open question for nonconvex functions. Recent results in \cite{kyrillidis2018run} suggest that AGD can achieve linear convergence rate for general nonconvex SDP problem. As compared to (as1), the initial condition for AGD's convergence  is updated to $\mathrm{Dist}(\u_0,\hat{\u}) \leq \frac{\rho_a}{\sqrt{\kappa}} \|\hat{\u}\|_2$ where $\rho_a$ is again a small constant. Clearly, this condition is more relaxed than (as1), as it increases the upper bound by a factor of $\sqrt{\kappa}$. In addition to the improved initial condition, our numerical tests have shown that the AGD-SE method will achieve faster convergence speed than FGD-SE, under the exactly same settings of initialization and step-size. 
\end{remark} 

\begin{algorithm}[tb!]
\caption{AGD-SE}
\label{alg:AGD}
\begin{algorithmic}[1]
\REQUIRE  Function $f$, rank $r$, maximum iteration number $K$.
\ENSURE  $\mathbf{u}$ and $\mathbf{V}=\mathbf{u}\mathbf{u}^\mathcal{H}$
\STATE   Initialize $\mathbf{u}_0\in \mathbb{C}^{n\times r}$ and set $\mathbf{V}_0=\mathbf{u}_0\mathbf{u}_0^\mathcal{H}$. 
\STATE Set the step-size $\eta$ as in \eqref{eq:eta}.
\STATE  Compute $\mathbf{u}_1=\mathbf{u}_0-\eta \nabla g(\mathbf{u}_0)$. 
\FOR{$k=1$ to $K$}
\STATE Update $\u^+$ as \eqref{eq:AGD predict} and  $\mathbf{u}_{k+1}=\mathbf{u}^+-\eta \nabla g(\mathbf{u}^+) $
\ENDFOR 
\RETURN $\mathbf{u}_K$ and $\mathbf{V}=\mathbf{u}_K\mathbf{u}_K^\mathcal{H}$
\end{algorithmic}
\end{algorithm}



\section{Practical Extensions for GD-SE Solvers}
\label{section4}

In addition to convergence guarantees and speed, it is truly important to consider practical issues in SE, such as i) robustness to bad data (outliers), ii) incorporation of additional meter types, and iii) multi-area implementation. Traditionally, outliers arise in SE due to data contamination, meter failure and synchronization issues  \cite{monticelli2000electric,zhao2018statistical,wang2019robust2}. More recently,  malicious cyber attacks \cite{liu2011false} and topology errors \cite{weng2015convexification} can also contribute to SE outliers. Meanwhile, recent development of phasor measurement units (PMUs)  motivates us to expand the SE solvers to incorporate synchrophasor data as well. There is also increasing trend to consider multi-area SE among different control centers \cite{zhu2014power,madani2017low,kekatos2013distributed}. All these practical concerns have been shown to potentially worsen the convergence issue of GN-SE method; see e.g., \cite{zhao2018statistical,weng2015convexification,zhu2014power}.  Due to page limit, we focus on discussing the first two  extensions  for GD-SE. Multi-area (Distributed) GD-SE can be developed using the framework of distributed linear SE method such as \cite{kekatos2013distributed}.

\subsection{Robust GD-SE against Outliers}

One popular approach to tackle the presence of bad data is to introduce  a sparse vector $\boldsymbol{\tau}\in \mathbb{R}^L$ with $\tau_\ell \neq 0$ indicating an outlier entry. This way, the robust (R)SDP-SE problem can be formulated as \cite{kekatos2017psse} 
\begin{subequations}\label{eq:Robust_SE} 
\begin{align}
\min_{\mathbf{V},\boldsymbol{\tau}} & ~f' (\mathbf{V},\boldsymbol{\tau})
= \sum_{\ell=1}^L \frac{1}{2}\big \{[z_\ell -\mathrm{Tr}(\mathbf{H}_\ell \mathbf{V})] -\tau_\ell \big \}^2 \label{eq:Robust_o}\\
\mathrm{s.t.} &~~ \mathbf{V} \succeq \mathbf{0},\: \|\boldsymbol{\tau}\|_0 \leq \rho L \label{eq:Robust_c}
\end{align} 
\end{subequations}
with the pseudo-norm $\|\boldsymbol{\tau}\|_0=\sum_{\ell} \mathds{1} \{\tau_\ell \ne 0\}$ as the number of nonzero entries and $\rho$ is the given fraction of outliers. Hence, the constraint \eqref{eq:Robust_c} ensures that $\boldsymbol{\tau}$ is sparse with at most $\rho L$ non-zero entries. 

To generalize the GD-SE methods to  the RSDP-SE problem \eqref{eq:Robust_SE}, we consider the nonconvex counterpart as $g'(\u,\boldsymbol{\tau}):=f'(\u\u^\mathcal{H},\boldsymbol{\tau})$. Note that the GD updates now have to satisfy the additional sparsity constraint  \eqref{eq:Robust_c}. To this end, we adopt the idea of hard thresholding \cite{chen2017robust,li2019nonconvex} to obtain the \textit{truncated} gradient by removing the data samples with  significantly high mismatch error. For a given integer $\gamma\leq L$, define the hard thresholding operator $\mathcal{D}_\gamma$  as
\begin{align}\label{hard_thr} 
\big [\mathcal{D}_\gamma(\boldsymbol{\chi})\big ]_\ell := \left\{ {\begin{array}{*{20}{l}}
  {\chi_\ell, \qquad \mathrm{if}\:|\chi_\ell| \geq |\boldsymbol{\chi}^{(\gamma)}|,}\\ 
  {0, \qquad \;\: \mathrm{otherwise},} 
\end{array}} \right. 
\end{align} 
where $|\boldsymbol{\chi}^{(\gamma)}|$ denotes the entry in $\boldsymbol{\chi}\in \mathbb{R}^L$ with the $\gamma$-largest absolute value. Equivalently, $\mathcal{D}$ throws away the $\gamma$-largest values of $\boldsymbol{\chi}$. For any given $\u$, one can use the thresholding operator on the instantaneous error mismatch, namely $\chi_\ell=z_\ell -\u^\mathcal{H}\mathbf{H}_\ell \u$.  Hence, by setting the outlier indicator $\boldsymbol{\tau}=\mathcal{D}_{\rho L}(\boldsymbol{\chi})$, the former always satisfies the sparsity constraint, and  
the gradient of $g'$ at $\u$  becomes
\begin{align} 
\nabla_\u ~g'(\mathbf{u},\boldsymbol{\tau}) & = 2 \nabla_\u f'(\u\u^\mathcal{H},\boldsymbol{\tau}) \times \u \nonumber\\
&=
\sum_{\ell=1}^L 2 \big (\u^\mathcal{H}\mathbf{H}_\ell \u - z_\ell +\tau_{\ell} \big ) \mathbf{H}_\ell \mathbf{u}. \label{robust-gradient}
\end{align} 
Since $\boldsymbol\tau$ is the hard thresholded output of $\boldsymbol{\chi}$, the summation in \eqref{robust-gradient} only takes  $(L-\rho L)$ measurements of  smaller mismatch errors and rules out the rest of higher mismatch errors.

\begin{algorithm}[t!]
\caption{RFGD-SE and RAGD-SE}
\label{alg:RAGD}
\begin{algorithmic}[1]
\REQUIRE  Function $f$, rank $r$, maximum  iteration number $K$, and outlier fraction $\rho$.
\ENSURE  $\mathbf{u}$ and $\mathbf{V}=\mathbf{u}\mathbf{u}^\mathcal{H}$

\STATE   Initialize $\mathbf{u}_0\in \mathbb{C}^{n\times r}$ and set $\mathbf{V}_0=\mathbf{u}_0\mathbf{u}_0^\mathcal{H}$. 

\STATE  Compute the current error $\boldsymbol\chi_0$ and threshold it to $\boldsymbol{\tau}_0$. 

\STATE  Update $\mathbf{u}_1=\mathbf{u}_0-\eta \nabla_\u ~g'(\mathbf{u}_0,\boldsymbol{\tau}_0)$. 
\STATE Set the step-size $\eta$ as in \eqref{eq:eta}.
\FOR{$k=1$ to $K$}
\IF{RAGD updates}
\STATE Compute $\mathbf{u}^+=\mathbf{u}_k+\textstyle(\frac{k-2}{k+1})(\mathbf{u}_k-\mathbf{u}_{k-1})$
\ELSIF{RFGD updates}
\STATE Set $\u^+=\u_k$
\ENDIF
\STATE Compute the current error $\boldsymbol{\chi}_k=\mathbf{z}-\mathcal{H}(\u^+(\u^+)^\mathcal{H})$.\\
\STATE Threshold the error to 
$\boldsymbol{\tau}_k=\mathcal{D}_{\rho L}(\boldsymbol{\chi}_k)$.\\
\STATE Update 
$\mathbf{u}_{k+1}=\mathbf{u}^+-\eta \nabla_\u ~g'(\mathbf{u}^+,\boldsymbol{\tau}_k) .$
\ENDFOR 
\RETURN $\mathbf{u}_K$ and $\mathbf{V}=\mathbf{u}_K\mathbf{u}_K^\mathcal{H}$
\end{algorithmic}
\end{algorithm}

Using the truncated gradient in \eqref{robust-gradient}, we can develop the robust (R)FGD and (R)AGD methods for SE, as tabulated in Algorithm \ref{alg:RAGD}. The two algorithms are jointly presented as they are only different in the gradient updates, on either $\u^+$ or $\u_k$. Other settings such as initialization and step-size follow from the original FGD/AGD method. 

\begin{remark}
{\emph (Performance of Robust GD-SE) } The GD based updates using truncated gradient in \eqref{robust-gradient} has been proposed for solving nonconvex optimization problems under outliers, such as the robust phase retrieval problem \cite{chen2017robust,zhang2018median}, and more recently the robust matrix factorization problem \cite{li2019nonconvex}. Results therein suggest the truncated GD can achieve linear convergence rate with sufficiently good initialization, given that the original function is strongly convex and smooth. However, the convergence guarantee and rate of the truncated AGD updates are still an open question for nonconvex optimization. Our numerical tests have shown that the RAGD-SE method could achieve accelerated convergence performance than RFGD-SE, while both are shown to converge numerically.
\end{remark}

\subsection{Augmented GD-SE with PMU Data}

Compare to legacy quadratic measurement model of $\v$, PMUs provide synchronous phasor data that are linearly related to the state $\v$. If bus $n$ is equipped with a PMU, then its voltage phasor $V_n$ and the incident line current phasors $\{I_{nn'}\}$ are available with high accuracy. When there are sufficient PMU measurements making the system observable, SE will be non-iterative and fast thanks to the the linearity between the PMU measurements and the system states. Currently and in near future, power systems still have limited PMUs due to the high installation and networking costs. Hence,  SE must be performed using both the legacy meters and PMU data.

Let $\{\boldsymbol{\zeta}_n\}_{n \in \mathcal{P}}$ denote all the PMU measurements, where $\mathcal{P} \subseteq \mathcal{N}$ denotes the PMU-instrumented buses. The noisy PMU data at bus $n$ can be modeled as $\boldsymbol{\zeta}_n=\mathbf{\Phi}_n \v+\boldsymbol{\varepsilon}_n$, where $\mathbf{\Phi}_n$ denotes the measurement matrix constructed in accordance to the bus index $n$ and line parameters \cite{kekatos2012optimal} and $\boldsymbol{\varepsilon}_n$ the measurement noise assumed to be complex zero-mean Gaussian. 
Given both $\mathbf{z}$ and $\{\boldsymbol{\zeta}_n\}_{n \in \mathcal{P}}$, the joint SE problem aims to minimize the augmented error objective as 
\begin{align}\label{eq:PMU_wls}
\min_{\mathbf u\in\mathbb{C}^N} g^a(\u):=  g(\u) + \frac{1}{2} \sum_{n \in \mathcal{P}} \|\boldsymbol{\zeta}_n-\mathbf{\Phi}_n \u \|_2^2.
\end{align}
The PMU-augmented SE problem \eqref{eq:PMU_wls} is still nonconvex due to  $g(\u)$. 
Nonetheless, its gradient function can be formed by 
\begin{align}\label{eq:gradga}
\nabla g^a(\mathbf{u})= \nabla g(\u) + \sum_{n\in\mathcal{P}} \mathbf{\Phi}_n^\mathcal{H}( \mathbf{\Phi}_n \u-\boldsymbol{\zeta}_n).
\end{align}
Using this gradient function, both FGD-SE and AGD-SE methods can be extended to include PMU data. The new LS term from PMU data can be thought of as an additional regularization term on the original objective, which can improve the effectiveness of gradient update thanks to the accuracy of PMU data. Accordingly, the augmented FGD-SE or AGD-SE methods could accelerate their respective counterparts, as verified by our numerical results.   

\section{Numerical Results}
\label{section5}

The proposed FGD/AGD-SE methods and their extensions have been tested on a laptop with Intel\textsuperscript{\textregistered} CPU @ 2.2GHz (8GB RAM) in the MATLAB\textsuperscript{\textregistered} R2017a simulator. They are compared with the SDP-SE solutions using the MATLAB-based optimization modeling package \texttt{CVX} \cite{grant2008cvx} together with SeDuMi \cite{sturm1999using} solver. Three power transmission system test cases, namely, the IEEE 118-bus, 300-bus, and the synthetic ACTIVSg2000-bus \cite{birchfield2017grid} systems are used with the pertinent power flow solver and GN-SE iterations implemented by the MATLAB-based toolbox MATPOWER \cite{zimmerman2011matpower}. 
To generate the measurements, random Gaussian noise (in p.u.) is added to the power flow output, with $\sigma_\ell = 0.02$ at line flow meters, 0.04 at power injection meters, 0.004 at voltage meters, and 0.0004 at PMUs, respectively. Empirical root mean-square estimation error (RMSE)
$||\hat{\mathbf{v}}-\mathbf{v}||_2/\|\mathbf{v}\|_2$ is computed by averaging over 100 Monte-Carlo realizations. For each realization, the actual bus voltage magnitude (in p.u.) and phase angle (in rad) of each bus are uniformly distributed over $[0.95,1.05]$ and  $[-0.35\pi,0.35\pi]$, respectively. To construct the normalized systems, we follow the earlier approaches \cite{gol2014lav,wang2019robust1} scale with measurement as  $\{ \frac{z_\ell}{\|\mathbf{H}_\ell\|_F}, \frac{\mathbf{H}_\ell}{\|\mathbf{H}_\ell\|_F} \}_{\ell=1}^L$.

To initialize all iterative updates, we set the bus voltage magnitude to be its measured value, or 1 p.u.. The phase angles are initialized as the linear SE solutions based on dc power flow model. Generally the SDP-SE solution is not perfectly rank-one. Thus, the popular eigen-decomposition approach is used to retrieve the best rank-one approximation; see e.g., \cite{zhu2014power}. All estimated voltage vectors by GD-SE or SDP-SE will be further improved using the GN updates to reduce the optimality gap. Empirically, 3-4 GN iterations are sufficient for convergence, and thus this additional computational time is neglected for evaluating the runtime of SDP-/GD-SE methods.


\begin{figure*}
\centering
    \subfloat{\label{118_fig}\includegraphics[width=0.32\textwidth]{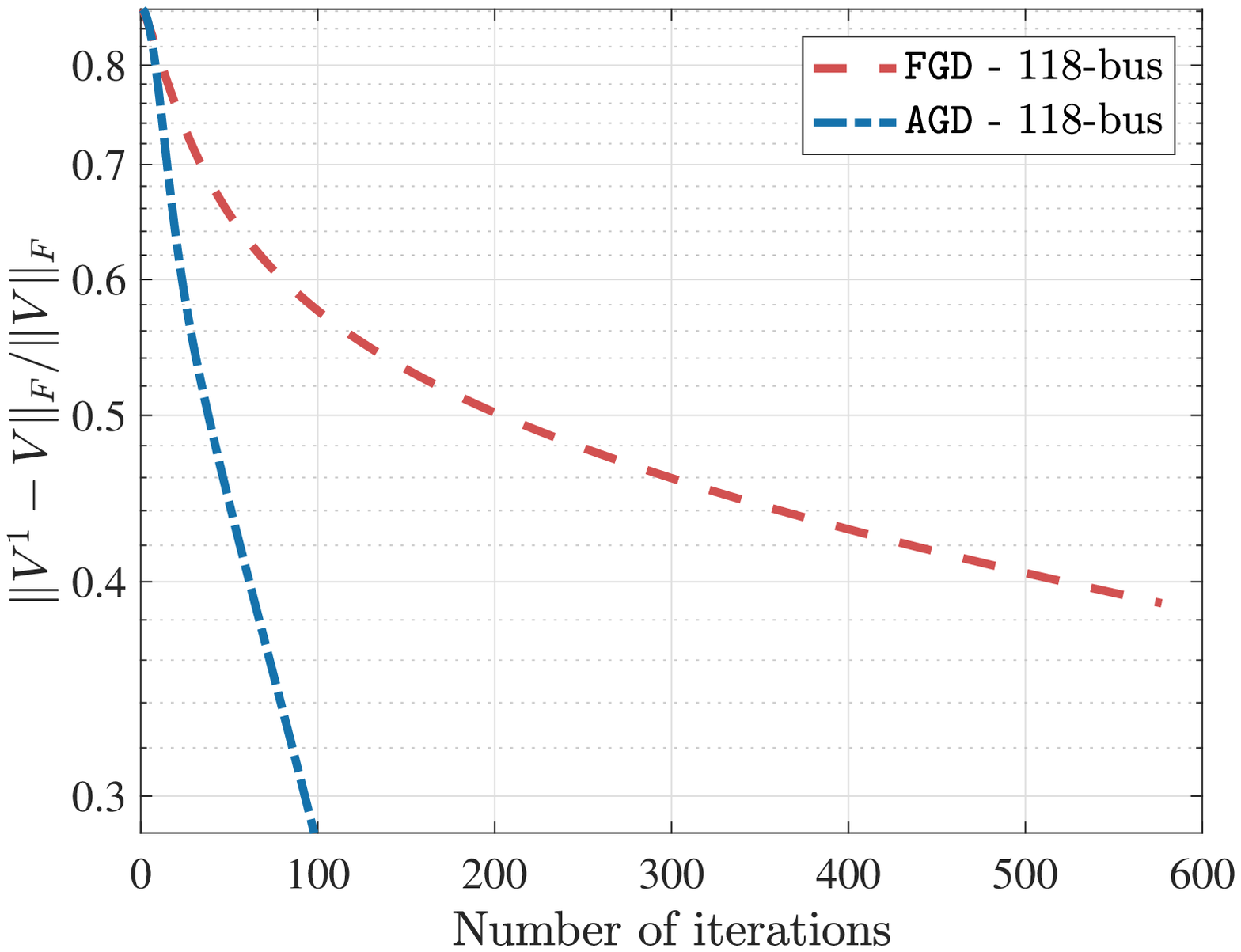}} 
    \subfloat{\label{300_fig}\includegraphics[width=0.32\textwidth]{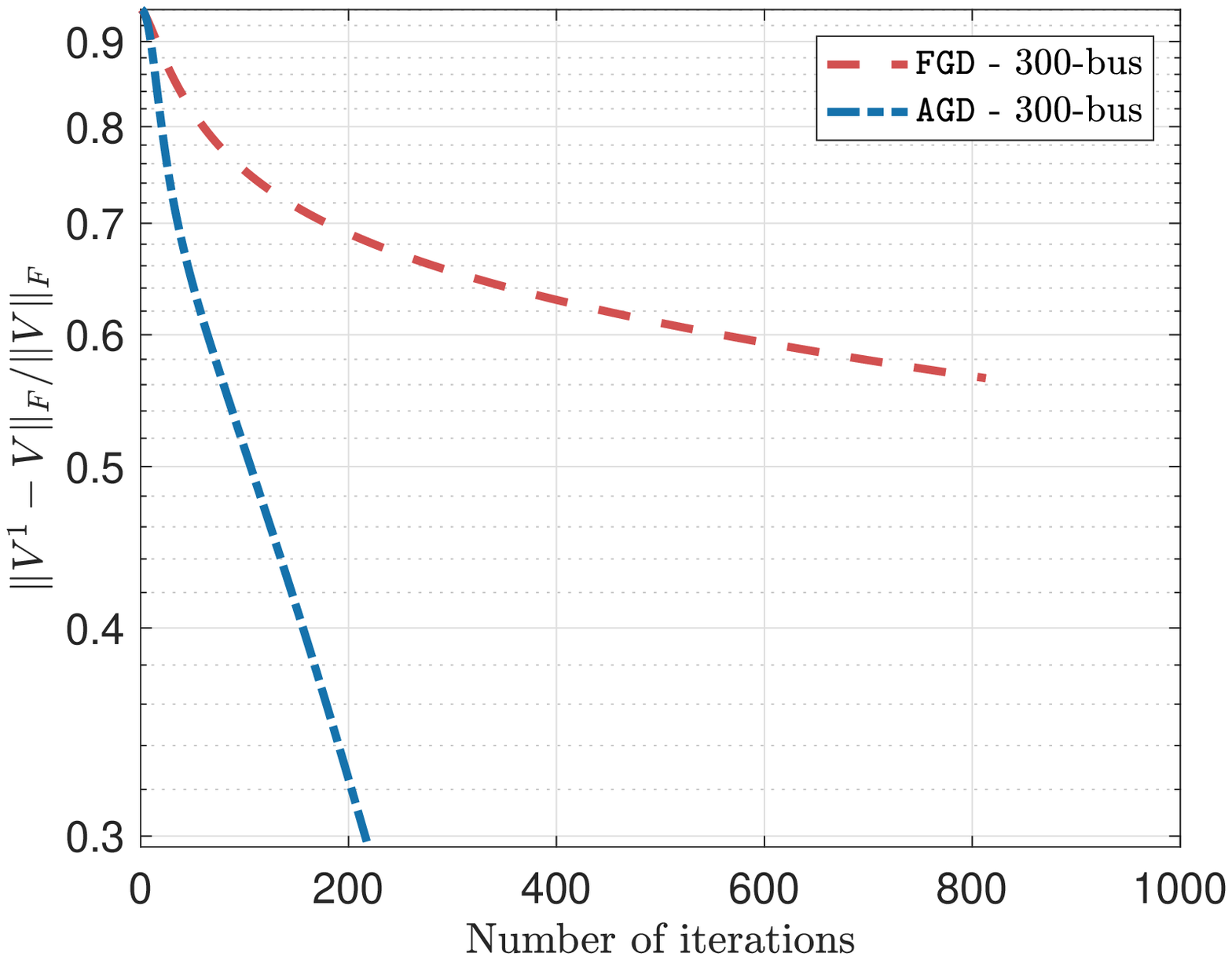}} 
    \subfloat
    {\label{2000_fig}\includegraphics[width=0.32\textwidth]{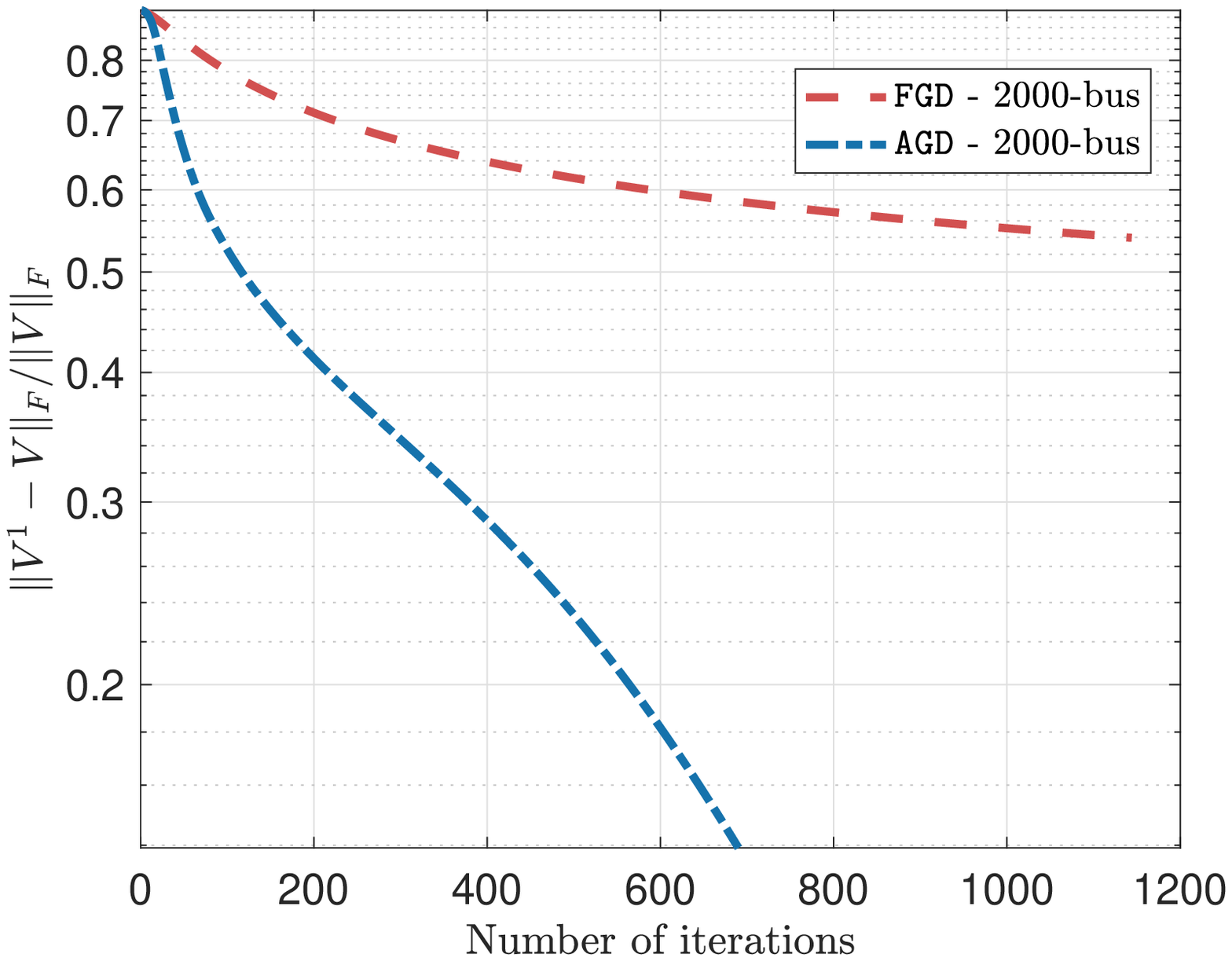}}
    \caption{Average iterative error from the {ground-truth $\V$} for the (left) 118-bus, (middle) 300-bus and (right) 2000-bus systems.}
    \vspace{-4mm}
    \label{fig:vsratios}
\end{figure*}

\vspace{3mm}

\textit{1) Test Case 1 on LS-SE Objective:} We first test all methods on the LS objective function \eqref{eq:gU}. The measurements include all bus voltage magnitudes, and all active and reactive line flows at the `from' end. The stopping criteria for GD-SE iterations  are based on consecutive change of the iterate and that of the objective value, while the stop criterion for GN is based on the first-order optimal condition. 

To demonstrate the performance of FGD and AGD in estimating {the ground-truth $\V$, the average iterative error $\|{\V}^1-\V\|_F/\|\V\|_F$,} where matrix $\V^1$ stands for the instantaneous matrix solution per iteration, is plotted in Fig. \ref{fig:vsratios}. Both GD methods use the same step-size, showing nearly linear convergence rate. Meanwhile, the AGD updates always outperform FGD in terms of the convergence speed, thanks to the improved condition number of AGD as mentioned in Sec. \ref{section3}. 


We also compare the SE performance achieved by GD-SE methods, to that by GN-SE and SDP-SE, as listed in Table \ref{table:GN/CVX/FGD estimator}, along with the percentage of convergence for the respective GN updates. As mentioned earlier, the SDP-SE and GD-SE solutions are used to initialize the GN updates (indicated by the -GN) for reduced optimality gap. Table \ref{table:GN/CVX/FGD estimator}  shows that these solutions achieve 100\% convergence rate, verifying their near-optimal performance. As a comparison, the GN-SE experiences divergence issues, even more seriously as the system sizes increases.  In addition, both FGD-GN and AGD-GN achieve the same SE performance as the benchmark SDP-GN. For the largest 2000-bus case, SDP-SE cannot be solved by the generic CVX solver in reasonable time, and thus its error performance is not reported. We can notice that FGD-/AGD-GN still demonstrate superior error performance for the 2000-bus case. 

\begin{table}
\caption{RMSE and GN Convergence Rate}
\label{table:GN/CVX/FGD estimator}
\centering  
\setlength{\tabcolsep}{0.5mm}{
\begin{tabular}{c|c|c|c}
\hline
SE Error   & 118-bus     & 300-bus      & 2000-bus       \\ \hline  
GN           &0.014 (93\%) &0.103 (76\%)  &0.256 (30\%) \\        
SDP-GN  &0.003 (100\%) &0.017 (100\%) &N/A (N/A)  \\
FGD-GN  &0.003 (100\%) &0.017 (100\%) &0.004 (100\%)\\
AGD-GN  &0.003 (100\%) &0.017 (100\%) &0.004 (100\%)\\
\hline
\end{tabular}}
\end{table}

\begin{table}
\caption{Average Runtime of All LS-SE Solvers}
\label{table:CVX/FGD time}
\centering  
\begin{tabular}{c|c|c|c}  
\hline
Time     &118-bus   & 300-bus   & 2000-bus \\ \hline  
GN       &0.124s   & 2.039s   & 151.801s \\            
SDP     &8.650s  & 98.557s  & N/A \\ 
FGD     &0.148s   & 0.749s   & 73.101s\\
AGD     &0.049s   & 0.277s   & 35.733s\\
\hline
\end{tabular}
\vspace{-2mm}
\end{table}

To better investigate the computational time improvement, we have listed the average runtime of all solution techniques in Table \ref{table:CVX/FGD time}. Compared to the SDP solver, both FGD and AGD scale nicely with the system size, especially for AGD thanks to the improved condition number. Since the SDP solution is not found for the 2000-bus case, its runtime is not reported. 
In contrast, AGD takes less than one minute to converge. Compared to the last two GD-based solutions, GN using the same initialization has shown much longer computational time, mainly because of its divergence issue. This comparison again verifies the improvement of gradient-based iterations, in approaching the globally optimal SE solutions. 

\vspace{3mm}

\textit{2) Test Case 2 on Robust SE:}
To generate outliers, five measurement meters are randomly picked in each Monte-Carlo run, with the corresponding data purposely manipulated to 5 times of actual value. 
To find the benchmark performance for the proposed RFGD/RAGD method, we consider a reformulated robust SDP-SE model (RSDP) by using the $l_1$-norm instead of $l_0$-norm term in \eqref{eq:Robust_SE} \cite{zheng2017distributed}. Specifically, we solve its Lagrangian form augmenting the objective function as ${\hat{\mathbf V}}= \arg \min_{\mathbf{V},\boldsymbol{\tau}}\: f' (\mathbf{V},\boldsymbol{\tau})+\lambda \|\boldsymbol{\tau}\|_1$, where $\lambda$ is a positive parameter. 
We also implement the classical robust least-absolute value (LAV) estimator \cite{monticelli2000electric}, which iteratively solves the linearized problem using the latest iterates.  Recently, the robust LAV objective has been solved using the so-termed stochastic prox-linear (SPL) method \cite{wang2019robust2}, which can accelerate the convergence speed by using a special linearization on the complex phasor representation.


\begin{table}[!t] 
\caption{RMSE and Outlier Identification Rate}
\label{table:robust tests}
\centering 
\setlength{\tabcolsep}{0.5mm}{ 
\begin{tabular}{c|c|c|c}  
\hline
SE      &118-bus   & 300-bus   & 2000-bus \\ \hline  
LAV			&0.207(30\%)  & 0.362(11\%)   & 1.245 (2\%)\\ 
SPL			&0.233(4\%)  & 0.358(3\%)   & 0.451 (2\%)\\  
RSDP      &0.028(60\%)  & 0.078(45\%)  & N/A (N/A) \\ 
RFGD      &0.027(67\%)  & 0.065(54\%)   & 0.029 (55\%)\\
RAGD      &0.016(73\%)  & 0.054(60\%)   & 0.018 (72\%)\\
\hline
\end{tabular}}
\end{table}

\begin{table}[!t] 
\caption{Average Runtime of Robust SE Solutions}
\label{table:robust time}
\centering  
\begin{tabular}{c|c|c|c}  
\hline
Time     &118-bus   & 300-bus   & 2000-bus \\ \hline  
LAV       &17.185s   & 42.623s   & 132.423s \\
SPL       &1.424s   & 4.131s   & 70.342s \\            
RSDP     &8.345s  &120.768s  & N/A \\ 
RFGD     &0.419s   & 1.776s   & 79.165s\\
RAGD     &0.226s   & 1.034s   & 42.188s\\
\hline
\end{tabular}
\vspace{-2mm}
\end{table}

\begin{figure}[t] 
\centering 
\includegraphics[trim = 88mm 0mm 40mm 0mm, clip,width=0.6\textwidth]{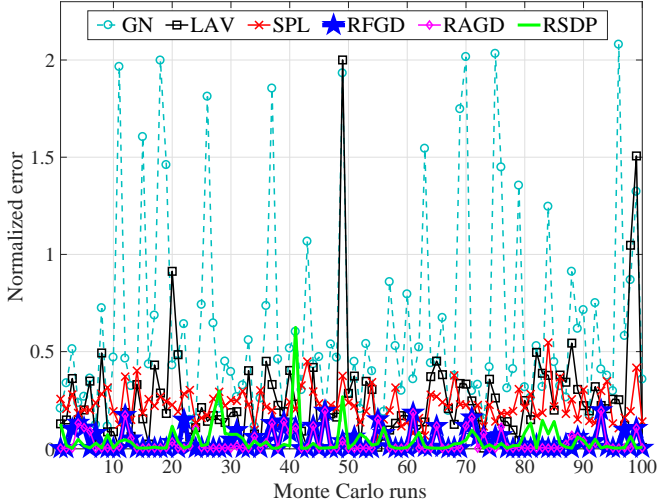} 
\caption{Estimation performance for the IEEE 118-bus system with 5 actual outliers.} 
\label{fig.robust compare} 
\end{figure}

We test all robust solvers, namely LAV, SPL, RSDP, RFGD/RAGD, on the three test cases. 
For simplicity, the initialization is set to be flat start. Again, the $l_1$-norm based RSDP is solved by CVX, where $\lambda$ is set based on the level of $3$ times noise standard deviation. As for RFGD/RAGD, the number of outliers is set to be 10 for the hard thresholding operation. Upon the convergence of all solution methods, 10 measurements with the largest deviation error will be identified as outliers and compared to the ground truth to compute the identification success rate. Afterwards, the other measurements together with the solution of $\u$ will be used to evoke the GN iterations to find the estimated voltage phasors and evaluate the SE error performance.

Table \ref{table:robust tests} lists the RMSE and success rate of identifying outliers for all robust SE methods, while Table \ref{table:robust time} provides the average runtime. Similar to Test Case 1, RFGD and RAGD consistently outperform the other robust solutions, in both the SE error and outlier identification rate, with similar computational time. Compared to Table \ref{table:GN/CVX/FGD estimator}, all the methods experience slightly larger RMSE because of the existence of outlier data. However, the performance of RFGD/RAGD is still elegant as they both can identify the majority of outliers. Interestingly for the robust case, RAGD at fastest runtime slightly outperforms RFGD in terms of error performance. Also, both of them can even improve the RSDP solution. This is perhaps because the robust GD methods perform outlier thresholding at every iteration and thus exclude outlier data more effectively.  Notably, as system size increases, the outlier identification rate becomes lower for both LAV and SPL methods. This points out the importance of having a well-designed robust SE algorithm for practical systems. 

Last, we have plotted the normalized error for each Monte-Carlo run of the 118-bus system in Fig. \ref{fig.robust compare}, including the LS-based GN solution. Evidently, all robust SE methods have improved over the GN-SE method due to the outlier data. The GD-based robust SE solutions consistently outperform other counterparts at reasonable computation time. 


\vspace{3mm}

\begin{table}[t] 
\caption{RMSE and Average Runtime of PMU-aided SE}
\label{table:PMU tests}
\centering  
\begin{tabular}{c|c|c|c}  
\hline
Time      		&118-bus   				& 300-bus   & 2000-bus \\ \hline  
GN$_p$        &0.012 (0.289s)   & 0.091 (7.837s)   & 0.182 (151.504s) \\   
SDP$_p$      &0.002 (8.415s)  & 0.015 (99.070s)  & N/A (N/A) \\ 
FGD$_p$      &0.002 (0.068s)   & 0.015 (0.469s)   & 0.003 (72.848s)\\
AGD$_p$      &0.002 (0.019s)   & 0.015 (0.114s)   & 0.003 (30.626s)\\
\hline
\end{tabular}
\vspace{-2mm}
\end{table}

\textit{3) Test Case 3 on PMU-aided SE:} We further evaluate the SE performance with the additional PMU data on the three test cases. All methods in Test Case 1 are considered, with their PMU-aided counterparts denoted by the subscript $_p$ here. For GN$_p$, the sequential approach of including PMU data in \cite{zhou2006alternative} is adopted. It entails two steps: i) the LS-based SE is performed first to process the legacy measurements, followed by ii) a post-processing step which only involves a linear models together with PMU data. Four buses are selected to equip with PMUs, namely $\{10, 12, 27, 15\}$, based on the PMU placement work in \cite{kekatos2013distributed}. {For these buses, the legacy meters are no longer included.}

Similarly, Table \ref{table:PMU tests} lists the SE error and average runtime for all PMU-augmented solutions. Compare to Table \ref{table:GN/CVX/FGD estimator}, the SE error has improved for all scenarios, thanks to the accurate PMU data. 
Both FGD$_p$ and AGD$_p$ still achieve the same estimation performance as SDP$_p$. Compared to the runtime in Table \ref{table:CVX/FGD time}, the two GD-based solutions are even faster with PMU data, thanks to the additional regularization that the latter provides to the objective function.

\section{Conclusions}
\label{section6}
This paper presented a gradient descent (GD) based framework for solving the nonconvex SE formulation, in order to 
accelerate the convex SDP-based SE for power system monitoring. The SDP formulation can offer near-optimality performance and improve the divergence issue of the iterative GN method. To tackle the high computational complexity of SDP-SE, we propose to adopt the factored (F)GD-update for the nonconvex objective on voltage phasor vector. Furthermore, the accelerated (A)GD-update is developed to improve the condition of the objective function. For FGD-SE, a linear convergence rate is guaranteed based on the analysis of the strong convexity and smoothness of the LS objective, whereas simiar result also holds for AGD-SE. Both proposed FGD-/AGD-SE methods can be extended to include practical scenarios of measurement outliers and PMU data. Extensive numerical comparisons have demonstrated the near-optimal error performance of the proposed approaches, while greatly reducing the computation time. 

Interesting future research directions open up, including the convergence guarantee and rate  of robust AGD-SE method. Moreover, we are interested to pursue the constrained SDP extension for SE problem under voltage limits  or even optimal power flow problem.

\vspace{-3mm}
\begin{appendix} 
\subsection{Calculation of Complex Gradient of a Real Function }
\label{Appendix:gradient}
Consider a real function $g$ with the complex $\u$ input, the complex gradient direction is given by \cite[Ch. 4]{petersen2008matrix}
\begin{align}
\nabla g(\u)=\frac{\partial g(\u)}{\partial \Re (\u)}+\mathrm{j}\frac{\partial g(\u)}{\partial \Im (\u)}. 
\end{align}
For the nonconvex SE objective $g$ in \eqref{eq:gU}, we can use the chain-rule to find
\begin{align}\label{gradient_g_part}
\nabla g(\u)=\sum\nolimits_{\ell=1}^L (\mathbf{u}^\mathcal{H} \mathbf{H}_\ell \mathbf{u}-z_\ell) \nabla (\u^\mathcal{H}\mathbf{H}_\ell \mathbf{u}).
\end{align}


To find the second-term in \eqref{gradient_g_part}, let $\u=\mathbf{r}+\mathrm{j}\mathbf{x}$ and $\mathbf{H}_\ell=\mathbf{R}_\ell+\mathrm{j}\mathbf{X}_\ell$, and thus the real function $(\u^\mathcal{H}\mathbf{H}_\ell \mathbf{u}) =\text{Tr}[\mathbf{R}_\ell (\mathbf{r}\mathbf{r}^\mathcal{T}+\mathbf{x}\mathbf{x}^\mathcal{T})-\mathbf{X}_\ell(\mathbf{x}\mathbf{r}^\mathcal{T}-\mathbf{r}\mathbf{x}^\mathcal{T})]$. 
Thus, we have
\begin{align}
\frac{\partial (\u^\mathcal{H}\mathbf{H}_\ell \mathbf{u})}{\partial \mathbf{r}}\mathop =\limits^{(i)} 2\mathbf{R}_\ell\mathbf{r}-\mathbf{X}_\ell\mathbf{x}+\mathbf{X}_\ell^\mathcal{T}\mathbf{x} 
\mathop =\limits^{(ii)} 2\mathbf{R}_\ell\mathbf{r}-2\mathbf{X}_\ell\mathbf{x} \nonumber 
\end{align}
where $(i)$ is due to the derivative of trace, and $(ii)$ follows from $\mathbf{X}_\ell^\mathcal{T}=-\mathbf{X}_\ell$ as $\mathbf{H}_\ell$ is Hermitian. Simiarly, we can find the partial derivative with respect to $\mathbf{x}$ and thus the complex gradient $\nabla (\u^\mathcal{H}\mathbf{H}_\ell \mathbf{u})= (2\mathbf{R}_\ell\mathbf{r}-2\mathbf{X}_\ell\mathbf{x})+ \mathrm{j}(2\mathbf{R}_\ell\mathbf{x}+2\mathbf{X}_\ell\mathbf{r})$, which is exactly equal to the complex product of $2\mathbf{H}_\ell\u$. This completes the proof of \eqref{eq:gradient}.


\subsection{Analysis of the SDP-SE Objective Function $f$}
\label{Appendix:m and M}
Consider any $\mathbf V \in \mathcal V$, $\mathcal{H} (\V)$ is the vector of error-free measurements. 
Thus, we have 
\begin{align}\label{eq:A calculation}
\|\mathcal{H}(\mathbf{V})\|_2^2 =& \sum\limits_{n \in \mathcal{N}_V} |{V}_n|^4+\sum\limits_{(n,n') \in \mathcal{E}_S} {P}_{n n'}^2 +\sum\limits_{(n,n') \in \mathcal{E}_S} {Q}_{n n'}^2 \nonumber\\
+ &\textstyle \sum\limits_{n \in \mathcal{N}_S} {P}_n^2 + \sum\limits_{n \in \mathcal{N}_S} {Q}_n^2,
\end{align}
where the sets denote the corresponding meter locations. Using the power flow model, we can derive upper/lower bounds for each of the summand terms in \eqref{eq:A calculation}.
First, using the voltage limits in \eqref{eq:setV} we have
\begin{align}\label{eq:v bounds}
\underaccent{\bar}{V}^4 \leq |V_n|^4 \leq \bar{V}^4.
\end{align}
As for the line flows, a trivial lower bound is 0, with the upper bound given by 
\begin{align} \label{eq:Pf_upper bound}
&P_{n n'}^2+ Q_{n n'}^2 
=  |S_{n n'}|^2
=|V_n (y_{n n'}(V_n-V_{n'}))^\mathcal{H}|^2 \nonumber\\
\leq&  |V_n|^2  |y_{n n'}|^2 (|V_n|+|V_{n'}|)^2 \leq 4|y_{n n'}|^2\bar{V}^4, 
\end{align}
by applying 
the triangle inequality.
Similarly, the power injection is upper bounded by
\begin{align} \label{eq:Pn_upper bound}
&P_{n}^2+Q_{n}^2  
= |S_{n}|^2=|V_n i_{n}^\mathcal{H}|^2=|V_n (\sum_{\nu \in \mathcal{N}_n} y_{n \nu}V_{\nu})^\mathcal{H}|^2 \nonumber\\
 \leq & |V_n|^2 \Big(\sum_{(n,\nu) \in \mathcal{E}}|y_{n \nu}| \bar{V}\Big)^2 
\leq \Big(\sum_{(n,\nu) \in \mathcal{E}}|y_{n \nu}| \Big)^2\bar{V}^4. 
\end{align}

This way, we can find the upper/lower bounds for $\|\mathcal{H}(\mathbf{V})\|_2^2 $ to quantify the two positive parameters $m$ and $M$. Note that we use a trivial lower bound of 0 for power measurements. This bound can be further improved by assuming a minimal angular separation between the two end buses of any line or using the system's total power demand.  
\end{appendix}


\ifCLASSOPTIONcaptionsoff
\fi
\bibliographystyle{IEEEtran}   
\bibliography{ref}

\begin{thebibliography}{10}
\providecommand{\url}[1]{#1}
\csname url@samestyle\endcsname
\providecommand{\newblock}{\relax}
\providecommand{\bibinfo}[2]{#2}
\providecommand{\BIBentrySTDinterwordspacing}{\spaceskip=0pt\relax}
\providecommand{\BIBentryALTinterwordstretchfactor}{4}
\providecommand{\BIBentryALTinterwordspacing}{\spaceskip=\fontdimen2\font plus
\BIBentryALTinterwordstretchfactor\fontdimen3\font minus
  \fontdimen4\font\relax}
\providecommand{\BIBforeignlanguage}[2]{{%
\expandafter\ifx\csname l@#1\endcsname\relax
\typeout{** WARNING: IEEEtran.bst: No hyphenation pattern has been}%
\typeout{** loaded for the language `#1'. Using the pattern for}%
\typeout{** the default language instead.}%
\else
\language=\csname l@#1\endcsname
\fi
#2}}
\providecommand{\BIBdecl}{\relax}
\BIBdecl

\bibitem{wood2012power}
A.~J. Wood and B.~F. Wollenberg, \emph{Power generation, operation, and
  control}.\hskip 1em plus 0.5em minus 0.4em\relax John Wiley \& Sons, 2012.

\bibitem{giannakis2013monitoring}
G.~B. Giannakis, V.~Kekatos, N.~Gatsis, S.-J. Kim, H.~Zhu, and B.~F.
  Wollenberg, ``Monitoring and optimization for power grids: A signal
  processing perspective,'' \emph{IEEE Signal Processing Magazine}, vol.~30,
  no.~5, pp. 107--128, 2013.

\bibitem{monticelli2000electric}
A.~Monticelli, ``Electric power system state estimation,'' \emph{Proceedings of
  the IEEE}, vol.~88, no.~2, pp. 262--282, 2000.

\bibitem{zhao2018statistical}
J.~Zhao, L.~Mili, and R.~C. Pires, ``Statistical and numerical robust state
  estimator for heavily loaded power systems,'' \emph{IEEE Transactions on
  Power Systems}, vol.~33, no.~6, pp. 6904--6914, 2018.

\bibitem{zhu2014power}
H.~Zhu and G.~B. Giannakis, ``Power system nonlinear state estimation using
  distributed semidefinite programming,'' \emph{IEEE Journal of Selected Topics
  in Signal Processing}, vol.~8, no.~6, pp. 1039--1050, 2014.

\bibitem{weng2015convexification}
Y.~Weng, M.~D. Ili{\'c}, Q.~Li, and R.~Negi, ``Convexification of bad data and
  topology error detection and identification problems in ac electric power
  systems,'' \emph{IET Generation, Transmission \& Distribution}, vol.~9,
  no.~16, pp. 2760--2767, 2015.

\bibitem{zhang2017conic}
Y.~Zhang, R.~Madani, and J.~Lavaei, ``Conic relaxations for power system state
  estimation with line measurements,'' \emph{IEEE Transactions on Control of
  Network Systems}, vol.~5, no.~3, pp. 1193--1205, 2017.

\bibitem{madani2016convexification}
R.~Madani, J.~Lavaei, and R.~Baldick, ``Convexification of power flow equations
  for power systems in presence of noisy measurements,'' \emph{IEEE
  Transactions on Automatic Control}, 2019.

\bibitem{kekatos2017psse}
V.~Kekatos, G.~Wang, H.~Zhu, and G.~B. Giannakis, ``{PSSE redux: Convex
  relaxation, decentralized, robust, and dynamic approaches},'' \emph{arXiv
  preprint arXiv:1708.03981}, 2017.

\bibitem{madani2017power}
R.~Madani, J.~Lavaei, R.~Baldick, and A.~Atamt{\"u}rk, ``Power system state
  estimation and bad data detection by means of conic relaxation,'' in
  \emph{Proc. 50th Hawaii Intl. Conf. on System Sciences}, 2017.

\bibitem{wang2019robust1}
G.~Wang, G.~B. Giannakis, and J.~Chen, ``Robust and scalable power system state
  estimation via composite optimization,'' \emph{IEEE Transactions on Smart
  Grid}, 2019.

\bibitem{madani2017low}
R.~Madani, A.~Kalbat, and J.~Lavaei, ``A low-complexity parallelizable
  numerical algorithm for sparse semidefinite programming,'' \emph{IEEE
  Transactions on Control of Network Systems}, 2017.

\bibitem{burer2003nonlinear}
S.~Burer and R.~D. Monteiro, ``A nonlinear programming algorithm for solving
  semidefinite programs via low-rank factorization,'' \emph{Mathematical
  Programming}, vol.~95, no.~2, pp. 329--357, 2003.

\bibitem{burer2005local}
------, ``Local minima and convergence in low-rank semidefinite programming,''
  \emph{Mathematical Programming}, vol. 103, no.~3, pp. 427--444, 2005.

\bibitem{bhojanapalli2016dropping}
S.~Bhojanapalli, A.~Kyrillidis, and S.~Sanghavi, ``Dropping convexity for
  faster semi-definite optimization,'' in \emph{Conference on Learning Theory},
  2016, pp. 530--582.

\bibitem{nesterov1983method}
Y.~Nesterov, ``A method of solving a convex programming problem with
  convergence rate o (1/k2),'' in \emph{Soviet Mathematics Doklady}, vol.~27,
  no.~2, 1983, pp. 372--376.

\bibitem{kyrillidis2018run}
A.~Kyrillidis, S.~Ubaru, G.~Kollias, and K.~Bouchard, ``Run procrustes, run! on
  the convergence of accelerated procrustes flow,'' \emph{arXiv preprint
  arXiv:1806.00534}, 2018.

\bibitem{gomez2004power}
A.~Gomez-Exposito and A.~Abur, \emph{Power system state estimation: theory and
  implementation}.\hskip 1em plus 0.5em minus 0.4em\relax CRC press, 2004.

\bibitem{zhang2018spurious}
R.~Zhang, J.~Lavaei, and R.~Baldick, ``Spurious critical points in power system
  state estimation,'' in \emph{Proceedings of the 51st Hawaii International
  Conference on System Sciences}, 2018.

\bibitem{wang2019robust2}
G.~Wang, H.~Zhu, G.~B. Giannakis, and J.~Sun, ``Robust power system state
  estimation from rank-one measurements,'' \emph{IEEE Transactions on Control
  of Network Systems}, 2019.

\bibitem{sturm1999using}
J.~F. Sturm, ``Using sedumi 1.02, a matlab toolbox for optimization over
  symmetric cones,'' \emph{Optimization methods and software}, vol.~11, no.
  1-4, pp. 625--653, 1999.

\bibitem{boyd2004convex}
S.~Boyd and L.~Vandenberghe, \emph{Convex optimization}.\hskip 1em plus 0.5em
  minus 0.4em\relax Cambridge university press, 2004.

\bibitem{negahban2012restricted}
S.~Negahban and M.~J. Wainwright, ``Restricted strong convexity and weighted
  matrix completion: Optimal bounds with noise,'' \emph{Journal of Machine
  Learning Research}, vol.~13, no. May, pp. 1665--1697, 2012.

\bibitem{liu2011false}
Y.~Liu, P.~Ning, and M.~K. Reiter, ``False data injection attacks against state
  estimation in electric power grids,'' \emph{ACM Transactions on Information
  and System Security (TISSEC)}, vol.~14, no.~1, p.~13, 2011.

\bibitem{kekatos2013distributed}
V.~Kekatos and G.~B. Giannakis, ``Distributed robust power system state
  estimation,'' \emph{IEEE Transactions on Power Systems}, vol.~28, no.~2, pp.
  1617--1626, 2013.

\bibitem{chen2017robust}
J.~Chen, L.~Wang, X.~Zhang, and Q.~Gu, ``Robust wirtinger flow for phase
  retrieval with arbitrary corruption,'' \emph{arXiv: 1704.06256}, 2017.

\bibitem{li2019nonconvex}
\BIBentryALTinterwordspacing
Y.~Li, Y.~Chi, H.~Zhang, and Y.~Liang, ``{Non-convex low-rank matrix recovery
  with arbitrary outliers via median-truncated gradient descent},''
  \emph{Information and Inference: A Journal of the IMA}, 05 2019. [Online].
  Available: \url{https://doi.org/10.1093/imaiai/iaz009}
\BIBentrySTDinterwordspacing

\bibitem{zhang2018median}
H.~Zhang, Y.~Chi, and Y.~Liang, ``Median-truncated nonconvex approach for phase
  retrieval with outliers,'' \emph{IEEE Transactions on Information Theory},
  vol.~64, no.~11, pp. 7287--7310, 2018.

\bibitem{kekatos2012optimal}
V.~Kekatos, G.~B. Giannakis, and B.~Wollenberg, ``Optimal placement of phasor
  measurement units via convex relaxation,'' \emph{IEEE Transactions on power
  systems}, vol.~27, no.~3, pp. 1521--1530, 2012.

\bibitem{grant2008cvx}
M.~Grant, S.~Boyd, and Y.~Ye, ``Cvx: Matlab software for disciplined convex
  programming,'' 2008.

\bibitem{birchfield2017grid}
A.~B. Birchfield, T.~Xu, K.~M. Gegner, K.~S. Shetye, and T.~J. Overbye, ``Grid
  structural characteristics as validation criteria for synthetic networks,''
  \emph{IEEE Transactions on power systems}, vol.~32, no.~4, pp. 3258--3265,
  2017.

\bibitem{zimmerman2011matpower}
R.~D. Zimmerman, C.~E. Murillo-S{\'a}nchez, and R.~J. Thomas, ``Matpower:
  Steady-state operations, planning, and analysis tools for power systems
  research and education,'' \emph{IEEE Transactions on power systems}, vol.~26,
  no.~1, pp. 12--19, 2011.

\bibitem{gol2014lav}
M.~Gol and A.~Abur, ``Lav based robust state estimation for systems measured by
  pmus.'' \emph{IEEE Trans. Smart Grid}, vol.~5, no.~4, pp. 1808--1814, 2014.

\bibitem{zheng2017distributed}
W.~Zheng, W.~Wu, A.~Gomez-Exposito, B.~Zhang, and Y.~Guo, ``Distributed robust
  bilinear state estimation for power systems with nonlinear measurements,''
  \emph{IEEE Transactions on Power Systems}, vol.~32, no.~1, pp. 499--509,
  2017.

\bibitem{zhou2006alternative}
M.~Zhou, V.~A. Centeno, J.~S. Thorp, and A.~G. Phadke, ``An alternative for
  including phasor measurements in state estimators,'' \emph{IEEE transactions
  on power systems}, vol.~21, no.~4, pp. 1930--1937, 2006.

\bibitem{petersen2008matrix}
K.~B. Petersen, M.~S. Pedersen \emph{et~al.}, ``The matrix cookbook,''
  \emph{Technical University of Denmark}, vol.~7, no.~15, p. 510, 2008.

\end{thebibliography}





\end{document}